\documentclass[aps,prl,twocolumn,showpacs,preprintnumbers,superscriptaddress,amsmath,amssymb]{revtex4}
\usepackage{amsmath, amsthm,textcomp }
\usepackage{graphicx}
\usepackage{dcolumn}
\usepackage{bm}
\usepackage{amssymb}
\usepackage{natbib}
\usepackage{array}
\usepackage{multirow}

\makeatletter

\begin{document}

\title{Spectroscopic Properties of Nuclear Skyrme Energy Density Functionals}

\author{D. Tarpanov}
\affiliation{Institute of Theoretical Physics, Faculty of Physics, University of Warsaw,
Ho\.za 69, PL-00-681 Warsaw, Poland}
\affiliation{Institute for Nuclear Research and Nuclear Energy, 1784 Sofia,
Bulgaria}

\author{J. Dobaczewski}
\affiliation{Institute of Theoretical Physics, Faculty of Physics, University of Warsaw,
Ho\.za 69, PL-00-681 Warsaw, Poland}
\affiliation{Department of Physics, P.O. Box 35 (YFL),
University of Jyv\"askyl\"a, FI-40014  Jyv\"askyl\"a, Finland}
\affiliation{Helsinki Institute of Physics, P.O. Box 64, FI-00014 Helsinki, Finland}

\author{J. Toivanen}
\affiliation{Department of Physics, P.O. Box 35 (YFL),
University of Jyv\"askyl\"a, FI-40014  Jyv\"askyl\"a, Finland}

\author{B.G. Carlsson}
\affiliation{Division of Mathematical Physics, LTH, Lund University, Post Office Box 118,
S-22100 Lund, Sweden}

\begin{abstract}
We address the question of how to improve the agreement between
theoretical nuclear single-particle energies (SPEs) and observations.
Empirically, in doubly magic nuclei, the SPEs can be deduced from
spectroscopic properties of odd nuclei that have one more, or one
less neutron or proton. Theoretically, bare SPEs, before being
confronted with observations, must be corrected for the effects of the
particle vibration coupling (PVC). In the present work, we determine
the PVC corrections in a fully self-consistent way. Then, we adjust
the SPEs, with PVC corrections included, to empirical data. In
this way, the agreement with observations, on average, improves;
nevertheless, large discrepancies still remain. We conclude that the
main source of disagreement is still in the underlying mean fields,
and not in including or neglecting the PVC corrections.
\end{abstract}

\maketitle

Finite many-fermion systems, such as quantum dots, ultracold Fermi
gases, atoms, or atomic nuclei, exhibit conspicuous shell effects.
These can easily be modeled within mean-field approaches, which
assume that fermions occupy single-particle states in a common
one-body potential. In nuclei, the ensuing shell effects are responsible not only
for sequences of excited states in odd nuclei, and for
their quadrupole or magnetic moments, but also for deformation
properties, including the fission phenomena, or detailed features of
rotational bands~\cite{(Boh69),(Boh75),(Bra72),(Szy83)}.

A precise description of nuclear spectroscopic properties, that is,
those pertaining to single-particle structures, is one of the most
important goals of theory. The theoretical approach that is particularly
well suited to describe these structures is the
energy-density-functional (EDF) formalism, wherein the Kohn-Sham
single-particle orbitals play an essential role. Although the Kohn-Sham
single-particle energies (SPEs), or bare SPEs, have, in principle,
only an auxiliary meaning, see, e.g., the recent analysis in
Ref.~\cite{(Dug12)}, in nuclei they do provide a fair description of
masses and excited states of odd nuclei. The question of quantitative
determination of many-body corrections to nuclear SPEs is a matter of
ongoing debate, see
Refs.~\cite{(Col10),[Bor10],(Lit11),[Ind12],(Cao14a),(Gne14)}.

In practice, all nuclear EDFs currently used in applications depend
on parameters or coupling constants adjusted to empirical
data~\cite{[Ben03],(Kor14)}. In addition, most of them were
constructed by adjusting bare SPEs to selected empirical information.
Therefore, it is not at all clear to what extend the many-body
corrections were, or were not, included in the EDFs' parameters, and
thus whether it is legitimate to add them {\it a posteriori}.

In the present Letter, we take up the challenge of adjusting EDFs'
parameters to empirical SPEs {\em after} having added many-body
corrections. This is certainly the right way of proceeding, which was
never tried up to now, and which allows us to study the interplay
between the mean-field and beyond-mean-field effects on the SPEs.

We determined the many-body corrections to SPEs within the standard
particle-vibration-coupling (PVC) model~\cite{(Boh69),[Ber80w],[Mah85],[Slu93],(Col10)},
which is based on coupling
particles and holes with the RPA phonons up to second order of
perturbation theory.
The calculations were performed in a fully
self-consistent way, that is, the same Skyrme EDF
parametrization was used to determine the ground states of even-even
nuclei, single-particle states, RPA phonons, and particle-phonon
vertices.
The PVC correction $\delta \epsilon_{i}$ to the SPE
$\epsilon_{i}$ of the $i$th state has the form~\cite{(Col10)}:
\begin{eqnarray}\label{PVC}
 \delta \epsilon_{i}&=&\displaystyle \frac{1}{2j_{i}+1}
\left(    \sum_{nJp} \frac{\left|\langle i||V||p,nJ\rangle\right|^2 }
   {\epsilon_{i}-\epsilon_{p}-\hbar \omega_{nJ}+i\eta} \right. \nonumber \\
&&\hspace*{12mm}+\left. \sum_{nJh} \frac{\left|\langle i||V||h,nJ\rangle\right|^2 }
   {\epsilon_{i}-\epsilon_{h}+\hbar \omega_{nJ}-i\eta} \right),
\end{eqnarray}
where $\langle i||V||p,nJ\rangle$ and $\langle i||V||h,nJ\rangle$ are, respectively,
the standard particle-phonon and hole-phonon vertex reduced matrix
elements. Similarly as in Ref.~\cite{(Col10)}, a small imaginary
parameter is added to the denominator with $\eta=0.05$\,MeV.

Let us briefly discuss the physical contents of the PVC
correction~(\ref{PVC}). A rigorous density functional theory formalism based on the
Hohenberg-Kohn~\cite{[Hoh64a]} and Kohn-Sham~\cite{[Koh65a]} theorems
stipulates that there exists an exact universal functional, which
should give the exact lowest energies (in each quantum number) of
even and odd nuclei, which can be directly compared with experimental
masses. Needless to say, such an exact functional is not
known. However, we know that when a phenomenological EDF is minimized
in even and odd nuclei, the resulting odd-even mass differences are
{\em not} equal to the Kohn-Sham energies, see the recent
Ref.~\cite{(Tar14)} for discussions and further references. Then, the
so-called polarization corrections to particle and hole SPEs are
equal to diagonal terms in Eq.~(\ref{PVC}), for $i=p$ and $i=h$,
respectively. Full PVC correction~(\ref{PVC}) can thus be regarded
as an approximate way to generalize our functional so as
to model the degrees of freedom associated with the mixing of the
odd particle with particle-vibration coupled states.

In the present work, we concentrate on presenting results obtained
for the bare and PVC-corrected SPEs. As discussed above, the
former ones do not have physical meaning; however, they provide us
with a simple illustration of one-body nuclear
properties, and we show them below as an important background that
facilitates communication and comparison of results.

As a baseline of our analysis, we used a set of five different Skyrme
EDF parametrizations, SAMi~\cite{(Roc12a)}, SLy5~\cite{(Cha98)},
SIII~\cite{[Bei75]}, SkM*~\cite{(Bar82)}, and SkP~\cite{(Dob84)},
which are characterized by quite different effective masses, ranging
from $m^*/m=0.675$ to 1. We carried out the calculations using the
spherical solver {\sc{HOSPHE}}~\cite{[Car10d],[Car13]}, in which the
determination of the PVC corrections was implemented~\cite{[Tar14a]}.
The mean-field, RPA, and PVC solutions were obtained with a
harmonic-oscillator basis using 15 oscillator shells (17 shells for
$^{208}$Pb).

We included effects of phonons with both parities and considered
multipolarities ranging from $J=0$ to 15, although only for phonons
up to $J=6$ we obtained a significant impact on the results. The PVC
corrections were determined in the single-particle and phonon spaces
restricted to below 15 and 30\,MeV, respectively. In addition, only
significantly collective phonons, that is, those contributing more
than 5\% to the non-energy-weighted sum rule of the given channel,
were taken into account~\cite{(Col10),[Ber80w]}. A detailed analysis
of numerical conditions and convergence will be presented in the
forthcoming publication~\cite{[Tar14a]}.

\begin{figure}
\includegraphics[width=\columnwidth]{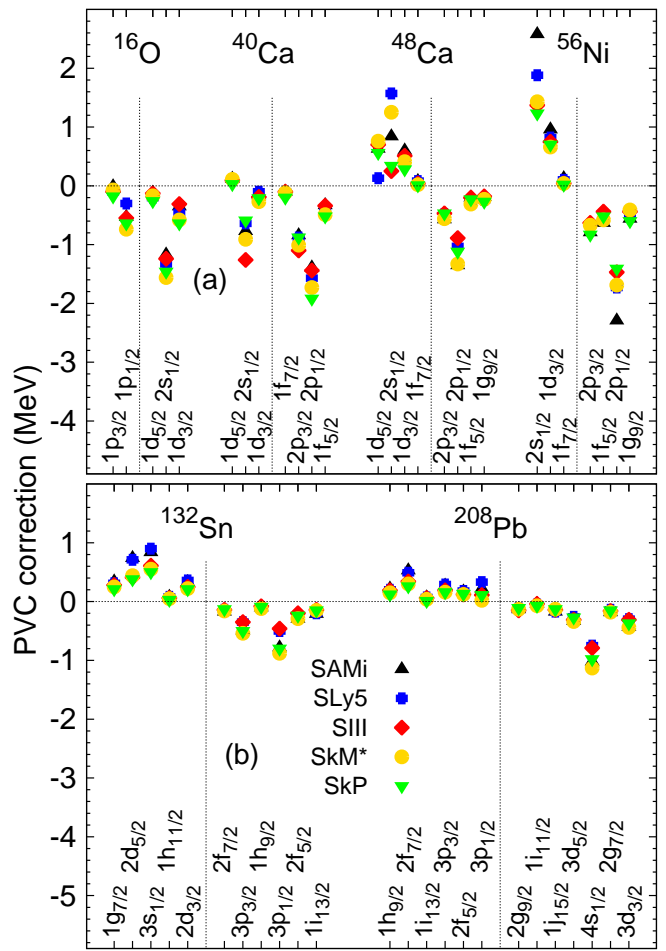}
\caption{(Color online) PVC corrections calculated for neutron SPEs in
six doubly magic nuclei and for five parametrizations of the Skyrme EDF.
For each nucleus, thin vertical lines separate hole and particle states.}
\label{Fig:1}
\end{figure}
\begin{figure}
\includegraphics[width=\columnwidth]{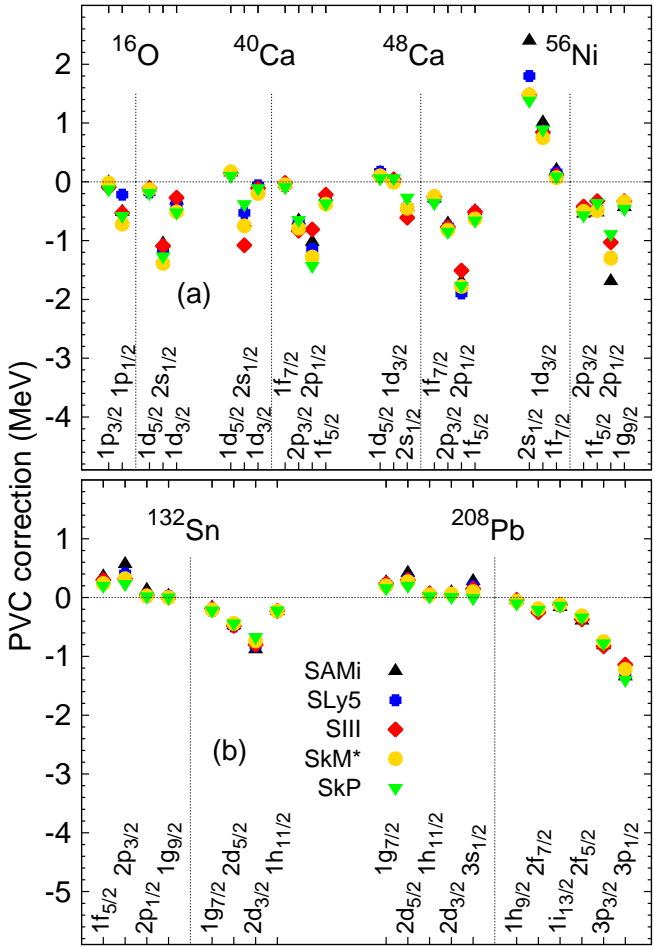}
\caption{(Color online) Same as in Fig.\ref{Fig:1} but for proton SPEs.}
\label{Fig:2}
\end{figure}
In Figs.~\ref{Fig:1} and~\ref{Fig:2}, we show values of the PVC
corrections calculated for neutron and proton SPEs, respectively, in
six doubly magic nuclei $^{16}$O, $^{40}$Ca, $^{48}$Ca, $^{56}$Ni,
$^{132}$Sn, and $^{208}$Pb. Those values are also tabulated in the
Supplemental Material~\cite{suppl}. We see that in some cases (e.g.,
$j=1/2$ states in $^{56}$Ni), the largest (smallest) PVC corrections
are obtained for the smallest (largest) effective masses; however,
the pattern of PVC corrections depends strikingly weakly on the EDF
parametrization.

Experimentally, the SPEs are not measurable quantities and they
cannot be defined in an entirely model-independent way~\cite{(Dug12)}.
They are
usually associated with masses and spectra of odd nuclei by
considering the so-called spectroscopic factors related to
probabilities of one-nucleon transfer reactions. Different analyses
of this type exist in the literature, and for the purpose of the
present study, we use those of Grawe {\it et
al.}~\cite{[Gra07a],(Fae13),[Gra14]} (data set A), Schwierz {\it et
al.}~\cite{[Sch07]} (data set B), and Porquet {\it  et
al.}~\cite{[Dud10],[Dud14],[Por14]} (data set C).
In addition, we also compare our results to two derived or reduced
data sets: (i) data set M, which contains average values of SPEs
simultaneously listed in data sets A, B, and C, provided the three
energies agree with the average values within 200\,keV, and (ii) data set
S, which contains a subset of data set B for spectroscopic factors
larger than 0.8. In this way, data set M contains SPEs, for which the
three evaluations agree best, and data set S contains those, which correspond
to least fragmented states. All data sets used in the present work are listed
in the Supplemental Material~\cite{suppl}.

In Fig.~\ref{Fig:4}, we show residuals of bare and PVC-corrected
SPEs, calculated with respect to empirical values of data set A. We
see that both bare and PVC-corrected SPEs poorly agree with data,
with deviations reaching up to around 4\,MeV. The distributions of
residuals are manifestly nonstatistical; hence, strong systematic
effects are still present~\cite{[Dob14a]}. Clearly, PVC corrections do
not improve the picture significantly. Simply adding the PVC corrections
to bare SPEs calculated for standard EDFs is insufficient, and the
readjustment of EDFs, as proposed in the present study, is mandatory.

\begin{figure}
\includegraphics[width=\columnwidth]{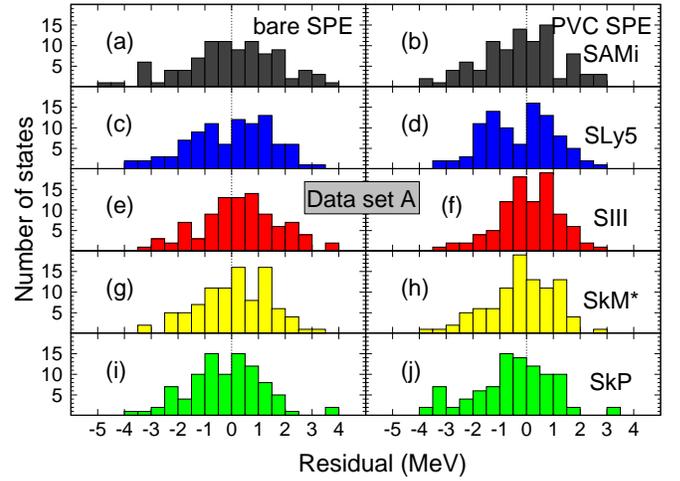}
\caption{(Color online) Distributions of residuals of the bare (left panels) and
PVC-corrected (right panels) SPEs, determined with respect to the
empirical values of data set A, and shown for the five
parametrizations of the Skyrme EDF.}
\label{Fig:4}
\end{figure}

To perform adjustments of the EDF coupling constants to empirical
SPEs, we follow the methodology of regression analysis, as it was applied in
Ref.~\cite{(Kor08)}. The method is based on the observation that
the standard Skyrme energy density, see e.g.~Refs.~\cite{[Eng75],[Per04]},
depends linearly on the 12 EDF coupling constants, $C_m$, $m=1,\ldots,12$.
In Ref.~\cite{(Kor08)}, it has been shown numerically that the above
linear dependence carries over to an approximate linear dependence of
bare SPEs on $C_m$. We are using this fact in
order to build the regression matrix
$I_{im}=\partial\epsilon_i/\partial{C_m}$, where the partial
derivatives are calculated using the finite-difference formula for
SPEs $\epsilon_i$ corresponding to coupling constants $C^0_m\pm{}d_m$,
perturbed by suitably small shifts $d_m$. In this way, we determine
the regression matrices for coupling constants $C^0_m$ corresponding
to five Skyrme EDF parametrizations considered in this study.

Using the regression matrices, and assuming that for reasonably small
changes of the coupling constants they do not significantly change,
one can fit the EDF coupling constants to the empirical SPEs. To
this end, one must solve the set of linear equations,
$r^0_i=\sum_m I_{im}\Delta C_m$,
where $r^0_i=\epsilon^0_i-\epsilon^{\text{exp}}_i$
are residuals of SPEs calculated for a given Skyrme EDF and
$\Delta{}C_m$ are corrections to coupling constants.
Since the numbers of empirical SPEs ($M=93$, 83, 78, 48, and 49 for data sets A, B, C, M, and S,
respectively) are larger than the number of coupling constants (12),
the best approximation is obtained within the standard
least-squares method, see, e.g.,~Ref.~\cite{[Bjo96]}, which minimizes
the rms deviation between the theory and experiment,
$\Delta \epsilon_{\text{rms}}= \left[ \tfrac{1}{M}  \sum_{i=1}^M (\epsilon_i
-\epsilon_i^{\text{exp}})^2 \right]^{1/2}$.

\begin{figure}
\includegraphics[width=\columnwidth]{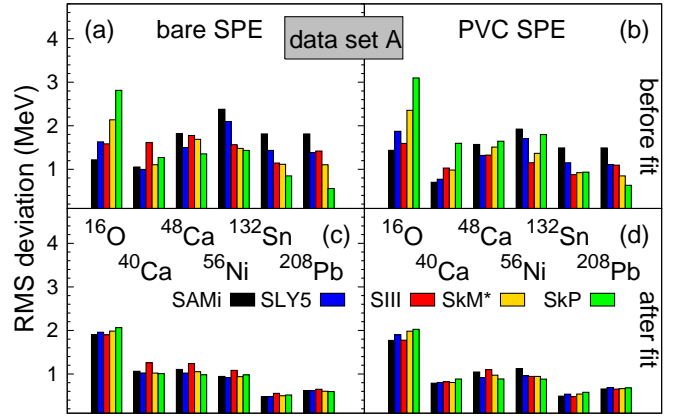}
\caption{(Color online) Left (right) panels: rms deviations between
the bare (PVC-corrected) SPEs and empirical data of set A. Upper and
lower panels show results obtained for standard Skyrme EDFs and for
refitted parametrizations, respectively. In all cases, partial
contributions obtained in six doubly magic nuclei are shown.}
\label{Fig:5}
\end{figure}
\begin{figure}[t]
\includegraphics[width=\columnwidth]{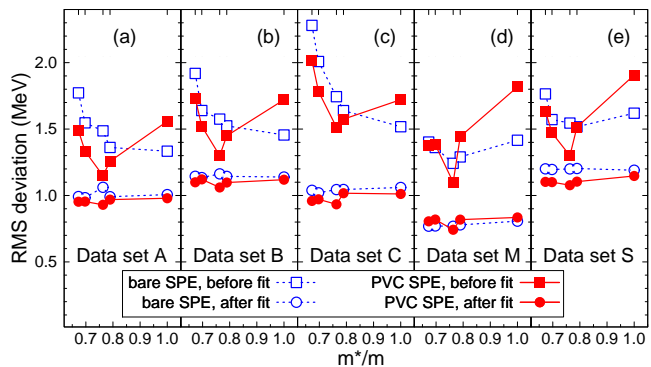}
\caption{(Color online) Open (full) symbols: rms deviations between
the bare (PVC-corrected) SPEs and empirical data of sets A, B, C, M, and S
(see text).
Results obtained for the five studied Skyrme EDFs (squares), and for the corresponding
refitted parametrizations (circles), are shown in function of the
effective mass corresponding to the original Skyrme EDF parametrizations.}
\label{Fig:7}
\end{figure}
Figs.~\ref{Fig:5} and~\ref{Fig:7} summarize results obtained after
the fits and compare them to those determined before the fits, that is,
to those corresponding to the original five Skyrme EDF parametrizations.
We note that the fitted values of SPEs were obtained directly from the regression analysis.
Fig.~\ref{Fig:5} shows partial rms deviations corresponding to the
six studied nuclei, whereas total rms deviations are shown in
Fig.~\ref{Fig:7}.

First, in Fig.~\ref{Fig:7} we see that the bare
SPEs obtained for standard Skyrme EDFs exhibit a conspicuous
effective-mass dependence, with those corresponding to $m^*/m=1$
being, on average, closest to data. This reflects the general
feature of adjusting EDFs to ground-state properties, whereupon one
systematically obtains the best fits for $m^*/m=1$~\cite{(Kor14)}.
In Fig.~\ref{Fig:5}(a) we also see that this trend is reversed in
$^{16}$O and absent in $^{40,48}$Ca, so the effect is clearly marked
only in heavy nuclei.

Second, in Fig.~\ref{Fig:7} we see that for the $m^*/m\le1$ EDFs, the
PVC-corrected SPEs agree (before the fit) slightly better with empirical SPEs
than the bare SPEs. The degree of improvement is systematic, but
small -- so small that in Fig.~\ref{Fig:4} it is not even really
visible. For the $m^*/m=1$ EDF, where already the bare SPEs agree
with data best, the PVC corrections lead to a deteriorated agreement.

Third, fits to bare and PVC-corrected SPEs, Figs.~\ref{Fig:5}(c,d)
and~\ref{Fig:7}, give rms deviations that are very weakly dependent
on variants of the Skyrme EDF. This indeed confirms the validity of
our regression analysis, because independently of the starting point,
linearity of the problem allows for bringing all optimized SPEs to
one common point. Note that these results indicate that the PVC corrections alone also depend
on the coupling constants approximately linearly. The remaining weak
dependence on the starting point may reflect possible small nonlinearities
as well as the fact the studied EDFs are defined with different powers
of the density dependence, which were not included in the regression analysis.
At the same time, we should bear in mind that the regression analysis
can be ill conditioned, cf.\ Ref.\ \cite{[Toi08]}, that is, while giving
robust values of the rms deviations, which are discussed in
this Letter, it may give poorly defined values of model
parameters.

Most important, in Fig.~\ref{Fig:7} we see that fits of
neither bare nor PVC-corrected SPEs can bring us below the glass
floor of about 1\,MeV of the total rms deviation. The independence of
this limit to the PVC corrections being included or not, shows that
they are not really giving us, on average, any better agreement with
empirical SPE's. Moreover, it also shows that the impact of the PVC
corrections on SPEs can be fairly well absorbed in the current
parametrization of the Skyrme EDF.

As seen in Fig.~\ref{Fig:7}, details of the comparison with
observations are still dependent on the way the empirical SPEs are
extracted from data. However, independently of which of the data sets A, B, or C
is used, the optimized results again do not go below the limit of about
1\,MeV rms deviation. In fact, the exact value of this limit
depends on the selection of the empirical data~\cite{suppl} -- considering only
those data points where data sets A, B, and C agree within $\pm$200\,keV (data set M),
the rms goes down to about 0.8\,MeV, and for those corresponding to spectroscopic
factors larger than 0.8 (data set S), it stays at
about 1.1\,MeV, where it also was for data set B.

In conclusion, the findings of the present work cast some doubts on
whether a rather poor agreement of bare SPEs, obtained within EDF
approaches, can be improved by taking into account PVC
corrections. We do not find any rationale in favor of adding these
corrections on top of results obtained for functionals that were
adjusted to observations without these corrections. In our study, we
obtained results for functionals refitted {\em after} adding the PVC
corrections. Although this was done within the approximation
based on linear regression, one could not improve the results below
the lower limit of about 1\,MeV rms deviation.

When fitting the coupling constants, we did not
consider any other observables, such as nuclear masses or radii.
Therefore, the obtained results simply illustrate the {\em maximum}
possible improvement that can potentially be obtained in the
description of the empirical SPEs. Taking into account other observables
can only worsen the results obtained for SPEs, which makes the conclusions
of our Letter even stronger. Certainly, PVC
effects are needed for a correct description of fragmented
single-particle strengths~\cite{(Miz12)}; however, for a detailed
description of values of SPEs, they do not lead to any dramatic improvement.

In our opinion, the burden of improving the current limited level of
agreement of nuclear spectroscopic properties with data is still on
the definition and form of the used EDFs~\cite{(Kor14)}, and not on
higher-order perturbative
corrections. Indeed, one cannot reasonably expect that many-body
corrections can compensate for the rather rudimentary forms of EDFs
currently in use. The search for better EDFs, which is currently
pursued in various directions, remains an important priority for the field.

\begin{acknowledgments}
Private communications from H. Grawe, J. Dudek, and M.-G. Porquet,
related to the empirical values of SPEs, are gratefully acknowledged.
We thank Gianluca Col\`o and Witek Nazarewicz for critical reading
of the manuscript and comments.
This work was supported in part by the THEXO JRA within the EU-FP7-IA
Project ENSAR (Grant No. 262010), by the Academy of Finland and University
of Jyv\"askyl\"a within the FIDIPRO program, by the Polish National Science
Center under Contract No. 2012/07/B/ST2/03907, by the Bulgarian Science
Fund under Contract No.\ NuPNET-SARFEN DNS7RP01/0003, and
by the ERANET-NuPNET grant SARFEN
of the Polish National Centre for Research and Development (NCBiR).
We acknowledge
the CSC-IT Center for Science Ltd., Finland, for the allocation of
computational resources.
B.G.C. thanks the Swedish Research Council (VR) for financial support.
\end{acknowledgments}

\bibliographystyle{apsrev}

\begin{thebibliography}{44}
\expandafter\ifx\csname natexlab\endcsname\relax\def\natexlab#1{#1}\fi
\expandafter\ifx\csname bibnamefont\endcsname\relax
  \def\bibnamefont#1{#1}\fi
\expandafter\ifx\csname bibfnamefont\endcsname\relax
  \def\bibfnamefont#1{#1}\fi
\expandafter\ifx\csname citenamefont\endcsname\relax
  \def\citenamefont#1{#1}\fi
\expandafter\ifx\csname url\endcsname\relax
  \def\url#1{\texttt{#1}}\fi
\expandafter\ifx\csname urlprefix\endcsname\relax\def\urlprefix{URL }\fi
\providecommand{\bibinfo}[2]{#2}
\providecommand{\eprint}[2][]{\url{#2}}

\bibitem[{\citenamefont{Bohr and Mottelson}(1969)}]{(Boh69)}
\bibinfo{author}{\bibfnamefont{A.}~\bibnamefont{Bohr}} \bibnamefont{and}
  \bibinfo{author}{\bibfnamefont{B.~R.} \bibnamefont{Mottelson}},
  \emph{\bibinfo{title}{Nuclear Structure, vol. I}} (\bibinfo{publisher}{W. A.
  Benjamin, New York}, \bibinfo{year}{1969}).

\bibitem[{\citenamefont{Bohr and Mottelson}(1975)}]{(Boh75)}
\bibinfo{author}{\bibfnamefont{A.}~\bibnamefont{Bohr}} \bibnamefont{and}
  \bibinfo{author}{\bibfnamefont{B.~R.} \bibnamefont{Mottelson}},
  \emph{\bibinfo{title}{Nuclear Structure, vol. II}} (\bibinfo{publisher}{W. A.
  Benjamin, Reading}, \bibinfo{year}{1975}).

\bibitem[{\citenamefont{Brack et~al.}(1972)\citenamefont{Brack, Damgaard,
  Jensen, Pauli, Strutinsky, and Wong}}]{(Bra72)}
\bibinfo{author}{\bibfnamefont{M.}~\bibnamefont{Brack}},
  \bibinfo{author}{\bibfnamefont{J.}~\bibnamefont{Damgaard}},
  \bibinfo{author}{\bibfnamefont{A.~S.} \bibnamefont{Jensen}},
  \bibinfo{author}{\bibfnamefont{H.~C.} \bibnamefont{Pauli}},
  \bibinfo{author}{\bibfnamefont{V.~M.} \bibnamefont{Strutinsky}},
  \bibnamefont{and} \bibinfo{author}{\bibfnamefont{C.~Y.} \bibnamefont{Wong}},
  \bibinfo{journal}{Rev. Mod. Phys.} \textbf{\bibinfo{volume}{44}},
  \bibinfo{pages}{320} (\bibinfo{year}{1972}).

\bibitem[{\citenamefont{Szyma{\'n}ski}(1983)}]{(Szy83)}
\bibinfo{author}{\bibfnamefont{Z.}~\bibnamefont{Szyma{\'n}ski}},
  \emph{\bibinfo{title}{Fast Nuclear Rotation}} (\bibinfo{publisher}{Clarendon
  Press}, \bibinfo{address}{Oxford}, \bibinfo{year}{1983}).

\bibitem[{\citenamefont{Duguet and Hagen}(2012)}]{(Dug12)}
\bibinfo{author}{\bibfnamefont{T.}~\bibnamefont{Duguet}} \bibnamefont{and}
  \bibinfo{author}{\bibfnamefont{G.}~\bibnamefont{Hagen}},
  \bibinfo{journal}{Phys. Rev. C} \textbf{\bibinfo{volume}{85}},
  \bibinfo{pages}{034330} (\bibinfo{year}{2012}).

\bibitem[{\citenamefont{Col\`o et~al.}(2010)\citenamefont{Col\`o, Sagawa, and
  Bortignon}}]{(Col10)}
\bibinfo{author}{\bibfnamefont{G.}~\bibnamefont{Col\`o}},
  \bibinfo{author}{\bibfnamefont{H.}~\bibnamefont{Sagawa}}, \bibnamefont{and}
  \bibinfo{author}{\bibfnamefont{P.~F.} \bibnamefont{Bortignon}},
  \bibinfo{journal}{Phys. Rev. C} \textbf{\bibinfo{volume}{82}},
  \bibinfo{pages}{064307} (\bibinfo{year}{2010}).

\bibitem[{\citenamefont{{P.F. Bortignon, G. Col\`o, and H. Sagawa, J. Phys.
  {\bf G37}, 064013 (2010)}}()}]{[Bor10]}
\bibinfo{author}{\bibnamefont{{P.F. Bortignon, G. Col\`o, and H. Sagawa, J.
  Phys. {\bf G37}, 064013 (2010)}}}.

\bibitem[{\citenamefont{Litvinova and Afanasjev}(2011)}]{(Lit11)}
\bibinfo{author}{\bibfnamefont{E.~V.} \bibnamefont{Litvinova}}
  \bibnamefont{and} \bibinfo{author}{\bibfnamefont{A.~V.}
  \bibnamefont{Afanasjev}}, \bibinfo{journal}{Phys. Rev. C}
  \textbf{\bibinfo{volume}{84}}, \bibinfo{pages}{014305}
  (\bibinfo{year}{2011}).

\bibitem[{\citenamefont{{A. Idini, F. Barranco, and E. Vigezzi, Phys. Rev. C
  {\bf 85}, 014331 (2012)}}()}]{[Ind12]}
\bibinfo{author}{\bibnamefont{{A. Idini, F. Barranco, and E. Vigezzi, Phys.
  Rev. C {\bf 85}, 014331 (2012)}}}.

\bibitem[{\citenamefont{Cao et~al.}(2014)\citenamefont{Cao, Col\`o, Sagawa, and
  Bortignon}}]{(Cao14a)}
\bibinfo{author}{\bibfnamefont{L.-G.} \bibnamefont{Cao}},
  \bibinfo{author}{\bibfnamefont{G.}~\bibnamefont{Col\`o}},
  \bibinfo{author}{\bibfnamefont{H.}~\bibnamefont{Sagawa}}, \bibnamefont{and}
  \bibinfo{author}{\bibfnamefont{P.~F.} \bibnamefont{Bortignon}},
  \bibinfo{journal}{Phys. Rev. C} \textbf{\bibinfo{volume}{89}},
  \bibinfo{pages}{044314} (\bibinfo{year}{2014}).

\bibitem[{\citenamefont{Gnezdilov et~al.}(2014)\citenamefont{Gnezdilov, Borzov,
  Saperstein, and Tolokonnikov}}]{(Gne14)}
\bibinfo{author}{\bibfnamefont{N.~V.} \bibnamefont{Gnezdilov}},
  \bibinfo{author}{\bibfnamefont{I.~N.} \bibnamefont{Borzov}},
  \bibinfo{author}{\bibfnamefont{E.~E.} \bibnamefont{Saperstein}},
  \bibnamefont{and} \bibinfo{author}{\bibfnamefont{S.~V.}
  \bibnamefont{Tolokonnikov}}, \bibinfo{journal}{Phys. Rev. C}
  \textbf{\bibinfo{volume}{89}}, \bibinfo{pages}{034304}
  (\bibinfo{year}{2014}).

\bibitem[{\citenamefont{{M. Bender, P.-H. Heenen, and P.-G. Reinhard, Rev. Mod.
  Phys. {\bf 75}, 121 (2003)}}()}]{[Ben03]}
\bibinfo{author}{\bibnamefont{{M. Bender, P.-H. Heenen, and P.-G. Reinhard,
  Rev. Mod. Phys. {\bf 75}, 121 (2003)}}}.

\bibitem[{\citenamefont{Kortelainen et~al.}(2014)\citenamefont{Kortelainen,
  McDonnell, Nazarewicz, Olsen, Reinhard, Sarich, Schunck, Wild, Davesne, Erler
  et~al.}}]{(Kor14)}
\bibinfo{author}{\bibfnamefont{M.}~\bibnamefont{Kortelainen}},
  \bibinfo{author}{\bibfnamefont{J.}~\bibnamefont{McDonnell}},
  \bibinfo{author}{\bibfnamefont{W.}~\bibnamefont{Nazarewicz}},
  \bibinfo{author}{\bibfnamefont{E.}~\bibnamefont{Olsen}},
  \bibinfo{author}{\bibfnamefont{P.-G.} \bibnamefont{Reinhard}},
  \bibinfo{author}{\bibfnamefont{J.}~\bibnamefont{Sarich}},
  \bibinfo{author}{\bibfnamefont{N.}~\bibnamefont{Schunck}},
  \bibinfo{author}{\bibfnamefont{S.~M.} \bibnamefont{Wild}},
  \bibinfo{author}{\bibfnamefont{D.}~\bibnamefont{Davesne}},
  \bibinfo{author}{\bibfnamefont{J.}~\bibnamefont{Erler}},
  \bibnamefont{et~al.}, \bibinfo{journal}{Phys. Rev. C}
  \textbf{\bibinfo{volume}{89}}, \bibinfo{pages}{054314}
  (\bibinfo{year}{2014}).

\bibitem[{\citenamefont{{V. Bernard and N. Van Giai, Nucl. Phys. {\bf A348}, 75
  (1980).}}()}]{[Ber80w]}
\bibinfo{author}{\bibnamefont{{V. Bernard and N. Van Giai, Nucl. Phys. {\bf
  A348}, 75 (1980).}}}

\bibitem[{\citenamefont{{C. Mahaux, P.F. Bortignon, R.A. Broglia, and C.H.
  Dasso, Phys. Rep. {\bf 120}, 1 (1985)}}()}]{[Mah85]}
\bibinfo{author}{\bibnamefont{{C. Mahaux, P.F. Bortignon, R.A. Broglia, and
  C.H. Dasso, Phys. Rep. {\bf 120}, 1 (1985)}}}.

\bibitem[{\citenamefont{{V. Van der Sluys, D. Van Neck, M. Waroquier, and J.
  Ryckebusch, Nucl. Phys. A {\bf 551}, 210 (1993)}}()}]{[Slu93]}
\bibinfo{author}{\bibnamefont{{V. Van der Sluys, D. Van Neck, M. Waroquier, and
  J. Ryckebusch, Nucl. Phys. A {\bf 551}, 210 (1993)}}}.

\bibitem[{\citenamefont{{P. Hohenberg and W. Kohn, Phys. Rev. {\bf 136}, B864
  (1964)}}()}]{[Hoh64a]}
\bibinfo{author}{\bibnamefont{{P. Hohenberg and W. Kohn, Phys. Rev. {\bf 136},
  B864 (1964)}}}.

\bibitem[{\citenamefont{{W. Kohn and L.J. Sham, Phys. Rev. {\bf 140}, A1133
  (1965)}}()}]{[Koh65a]}
\bibinfo{author}{\bibnamefont{{W. Kohn and L.J. Sham, Phys. Rev. {\bf 140},
  A1133 (1965)}}}.

\bibitem[{\citenamefont{Tarpanov et~al.}(2014)\citenamefont{Tarpanov, Toivanen,
  Dobaczewski, and Carlsson}}]{(Tar14)}
\bibinfo{author}{\bibfnamefont{D.}~\bibnamefont{Tarpanov}},
  \bibinfo{author}{\bibfnamefont{J.}~\bibnamefont{Toivanen}},
  \bibinfo{author}{\bibfnamefont{J.}~\bibnamefont{Dobaczewski}},
  \bibnamefont{and} \bibinfo{author}{\bibfnamefont{B.~G.}
  \bibnamefont{Carlsson}}, \bibinfo{journal}{Phys. Rev. C}
  \textbf{\bibinfo{volume}{89}}, \bibinfo{pages}{014307}
  (\bibinfo{year}{2014}).

\bibitem[{\citenamefont{Roca-Maza et~al.}(2012)\citenamefont{Roca-Maza, Col\`o,
  and Sagawa}}]{(Roc12a)}
\bibinfo{author}{\bibfnamefont{X.}~\bibnamefont{Roca-Maza}},
  \bibinfo{author}{\bibfnamefont{G.}~\bibnamefont{Col\`o}}, \bibnamefont{and}
  \bibinfo{author}{\bibfnamefont{H.}~\bibnamefont{Sagawa}},
  \bibinfo{journal}{Phys. Rev. C} \textbf{\bibinfo{volume}{86}},
  \bibinfo{pages}{031306} (\bibinfo{year}{2012}).

\bibitem[{\citenamefont{Chabanat et~al.}(1998)\citenamefont{Chabanat, Bonche,
  Haensel, Meyer, and Schaeffer}}]{(Cha98)}
\bibinfo{author}{\bibfnamefont{E.}~\bibnamefont{Chabanat}},
  \bibinfo{author}{\bibfnamefont{P.}~\bibnamefont{Bonche}},
  \bibinfo{author}{\bibfnamefont{P.}~\bibnamefont{Haensel}},
  \bibinfo{author}{\bibfnamefont{J.}~\bibnamefont{Meyer}}, \bibnamefont{and}
  \bibinfo{author}{\bibfnamefont{R.}~\bibnamefont{Schaeffer}},
  \bibinfo{journal}{Nucl. Phys. A} \textbf{\bibinfo{volume}{635}},
  \bibinfo{pages}{231} (\bibinfo{year}{1998}).

\bibitem[{\citenamefont{{M. Beiner, H. Flocard, N. Van Giai, and P. Quentin,
  Nucl. Phys. {\bf A238}, 29 (1975)}}()}]{[Bei75]}
\bibinfo{author}{\bibnamefont{{M. Beiner, H. Flocard, N. Van Giai, and P.
  Quentin, Nucl. Phys. {\bf A238}, 29 (1975)}}}.

\bibitem[{\citenamefont{Bartel et~al.}(1982)\citenamefont{Bartel, Quentin,
  Brack, Guet, and H{\aa}kansson}}]{(Bar82)}
\bibinfo{author}{\bibfnamefont{J.}~\bibnamefont{Bartel}},
  \bibinfo{author}{\bibfnamefont{P.}~\bibnamefont{Quentin}},
  \bibinfo{author}{\bibfnamefont{M.}~\bibnamefont{Brack}},
  \bibinfo{author}{\bibfnamefont{C.}~\bibnamefont{Guet}}, \bibnamefont{and}
  \bibinfo{author}{\bibfnamefont{H.-B.} \bibnamefont{H{\aa}kansson}},
  \bibinfo{journal}{Nucl.\ Phys.\ A} \textbf{\bibinfo{volume}{386}},
  \bibinfo{pages}{79} (\bibinfo{year}{1982}).

\bibitem[{\citenamefont{Dobaczewski et~al.}(1984)\citenamefont{Dobaczewski,
  Flocard, and Treiner}}]{(Dob84)}
\bibinfo{author}{\bibfnamefont{J.}~\bibnamefont{Dobaczewski}},
  \bibinfo{author}{\bibfnamefont{H.}~\bibnamefont{Flocard}}, \bibnamefont{and}
  \bibinfo{author}{\bibfnamefont{J.}~\bibnamefont{Treiner}},
  \bibinfo{journal}{Nucl. Phys.} \textbf{\bibinfo{volume}{A422}},
  \bibinfo{pages}{103} (\bibinfo{year}{1984}).

\bibitem[{\citenamefont{{B.G. Carlsson, J. Dobaczewski, J. Toivanen, and P.
  Vesel\'y, Comput. Phys. Commun. {\bf 181}, 1641 (2010)}}()}]{[Car10d]}
\bibinfo{author}{\bibnamefont{{B.G. Carlsson, J. Dobaczewski, J. Toivanen, and
  P. Vesel\'y, Comput. Phys. Commun. {\bf 181}, 1641 (2010)}}}.

\bibitem[{\citenamefont{{B.G. Carlsson, J. Toivanen, J. Dobaczewski, P.
  Vesel\'y, Y. Gao, and D. Ward (to be published)}}()}]{[Car13]}
\bibinfo{author}{\bibnamefont{{B.G. Carlsson, J. Toivanen, J. Dobaczewski, P.
  Vesel\'y, Y. Gao, and D. Ward (to be published)}}}.

\bibitem[{\citenamefont{{D. Tarpanov {\it et al.} (to be
  published)}}()}]{[Tar14a]}
\bibinfo{author}{\bibnamefont{{D. Tarpanov {\it et al.} (to be published)}}}.

\bibitem[{\citenamefont{{See Supplemental Material at [http://arxiv.org/e-print/1405.4823v2], which includes
  Refs.~\cite{[Bro88b],[ensdf]}, for results pertaining to data sets $A$, $B$,
  $C$, $M$, and $S$, and for a comparison with experimental levels in odd
  nuclei.}}()}]{suppl}
\bibinfo{author}{\bibnamefont{{See Supplemental Material at [http://arxiv.org/e-print/1405.4823v2], which
  includes Refs.~\cite{[Bro88b],[ensdf]}, for results pertaining to data sets
  $A$, $B$, $C$, $M$, and $S$, and for a comparison with experimental levels in
  odd nuclei.}}}

\bibitem[{\citenamefont{{B.A. Brown, Phys. Rev. C {\bf 58}, 220
  (1998)}}()}]{[Bro88b]}
\bibinfo{author}{\bibnamefont{{B.A. Brown, Phys. Rev. C {\bf 58}, 220
  (1998)}}}.

\bibitem[{\citenamefont{{Evaluated Nuclear Structure Data File,
  {http://www.nndc.bnl.gov/ensdf/}}}()}]{[ensdf]}
\bibinfo{author}{\bibnamefont{{Evaluated Nuclear Structure Data File,
  {http://www.nndc.bnl.gov/ensdf/}}}}.

\bibitem[{\citenamefont{{H. Grawe, K. Langanke, and G. Mart{\`{\i}}nez-Pinedo,
  Rep. Prog. Phys. {\bf 70}, 1525 (2007)}}()}]{[Gra07a]}
\bibinfo{author}{\bibnamefont{{H. Grawe, K. Langanke, and G.
  Mart{\`{\i}}nez-Pinedo, Rep. Prog. Phys. {\bf 70}, 1525 (2007)}}}.

\bibitem[{\citenamefont{Faestermann et~al.}(2013)\citenamefont{Faestermann,
  G\'orska, and Grawe}}]{(Fae13)}
\bibinfo{author}{\bibfnamefont{T.}~\bibnamefont{Faestermann}},
  \bibinfo{author}{\bibfnamefont{M.}~\bibnamefont{G\'orska}}, \bibnamefont{and}
  \bibinfo{author}{\bibfnamefont{H.}~\bibnamefont{Grawe}},
  \bibinfo{journal}{Progr. in Part. Nucl. Phys.} \textbf{\bibinfo{volume}{69}},
  \bibinfo{pages}{85} (\bibinfo{year}{2013}).

\bibitem[{\citenamefont{{H. Grawe, private communication}}()}]{[Gra14]}
\bibinfo{author}{\bibnamefont{{H. Grawe, private communication}}}.

\bibitem[{\citenamefont{{N. Schwierz, I. Wiedenhover, and A. Volya,
  arXiv:0709.3525}}()}]{[Sch07]}
\bibinfo{author}{\bibnamefont{{N. Schwierz, I. Wiedenhover, and A. Volya,
  arXiv:0709.3525}}}.

\bibitem[{\citenamefont{{J. Dudek, B. Szpak, M.-G. Porquet, H. Molique, K.
  Rybak, and B. Fornal, J. Phys. G {\bf 37}, 064031 (2010)}}()}]{[Dud10]}
\bibinfo{author}{\bibnamefont{{J. Dudek, B. Szpak, M.-G. Porquet, H. Molique,
  K. Rybak, and B. Fornal, J. Phys. G {\bf 37}, 064031 (2010)}}}.

\bibitem[{\citenamefont{{J. Dudek, B. Szpak, M.-G. Porquet, and B. Fornal,
  unpublished}}()}]{[Dud14]}
\bibinfo{author}{\bibnamefont{{J. Dudek, B. Szpak, M.-G. Porquet, and B.
  Fornal, unpublished}}}.

\bibitem[{\citenamefont{{M.-G. Porquet, private communication}}()}]{[Por14]}
\bibinfo{author}{\bibnamefont{{M.-G. Porquet, private communication}}}.

\bibitem[{\citenamefont{{J. Dobaczewski, W. Nazarewicz, and P.-G. Reinhard, J.
  Phys. G: Nucl. Part. Phys. {\bf 41}, 074001 (2014)}}()}]{[Dob14a]}
\bibinfo{author}{\bibnamefont{{J. Dobaczewski, W. Nazarewicz, and P.-G.
  Reinhard, J. Phys. G: Nucl. Part. Phys. {\bf 41}, 074001 (2014)}}}.

\bibitem[{\citenamefont{Kortelainen et~al.}(2008)\citenamefont{Kortelainen,
  Dobaczewski, Mizuyama, and Toivanen}}]{(Kor08)}
\bibinfo{author}{\bibfnamefont{M.}~\bibnamefont{Kortelainen}},
  \bibinfo{author}{\bibfnamefont{J.}~\bibnamefont{Dobaczewski}},
  \bibinfo{author}{\bibfnamefont{K.}~\bibnamefont{Mizuyama}}, \bibnamefont{and}
  \bibinfo{author}{\bibfnamefont{J.}~\bibnamefont{Toivanen}},
  \bibinfo{journal}{Phys. Rev. C} \textbf{\bibinfo{volume}{77}},
  \bibinfo{pages}{064307} (\bibinfo{year}{2008}).

\bibitem[{\citenamefont{{Y.M. Engel, D.M. Brink, K. Goeke, S.J. Krieger, and D.
  Vautherin, Nucl. Phys. {\bf A249}, 215 (1975)}}()}]{[Eng75]}
\bibinfo{author}{\bibnamefont{{Y.M. Engel, D.M. Brink, K. Goeke, S.J. Krieger,
  and D. Vautherin, Nucl. Phys. {\bf A249}, 215 (1975)}}}.

\bibitem[{\citenamefont{{E. Perli\'nska, S.G. Rohozi\'nski, J. Dobaczewski, and
  W. Nazarewicz, Phys. Rev. C {\bf 69}, 014316 (2004)}}()}]{[Per04]}
\bibinfo{author}{\bibnamefont{{E. Perli\'nska, S.G. Rohozi\'nski, J.
  Dobaczewski, and W. Nazarewicz, Phys. Rev. C {\bf 69}, 014316 (2004)}}}.

\bibitem[{\citenamefont{Bj{\"o}rck}(1996)}]{[Bjo96]}
\bibinfo{author}{\bibfnamefont{{\AA}.}~\bibnamefont{Bj{\"o}rck}},
  \emph{\bibinfo{title}{Numerical Methods for least squares problems}}
  (\bibinfo{publisher}{Society for Industrial and Applied Mathematics},
  \bibinfo{year}{1996}).

\bibitem[{\citenamefont{{J. Toivanen, J. Dobaczewski, M. Kortelainen, and K.
  Mizuyama, Phys. Rev. C {\bf 78}, 034306 (2008)}}()}]{[Toi08]}
\bibinfo{author}{\bibnamefont{{J. Toivanen, J. Dobaczewski, M. Kortelainen, and
  K. Mizuyama, Phys. Rev. C {\bf 78}, 034306 (2008)}}}.

\bibitem[{\citenamefont{Mizuyama et~al.}(2012)\citenamefont{Mizuyama, Col\`o,
  and Vigezzi}}]{(Miz12)}
\bibinfo{author}{\bibfnamefont{K.}~\bibnamefont{Mizuyama}},
  \bibinfo{author}{\bibfnamefont{G.}~\bibnamefont{Col\`o}}, \bibnamefont{and}
  \bibinfo{author}{\bibfnamefont{E.}~\bibnamefont{Vigezzi}},
  \bibinfo{journal}{Phys. Rev. C} \textbf{\bibinfo{volume}{86}},
  \bibinfo{eid}{034318} (\bibinfo{year}{2012}).

\end{thebibliography}

\end{document}